\documentclass{jpsj-suppl}
\usepackage{txfonts} 

\title{Competition between $\beta$-delayed proton  and $\beta$-delayed $\gamma$ decay of the exotic $T_z$ = -2 nucleus $^{56}$Zn and fragmentation of the IAS}

\author{ B. \textsc{Rubio}$^{1}$,  S. E. A. \textsc{Orrigo}$^{1}$, Y. \textsc{Fujita}$^{2,3}$, B. \textsc{Blank}$^{4}$, W. \textsc{Gelletly}$^{5}$, J. \textsc{Agramunt}$^{1}$, A. \textsc{Algora}$^{1}$, P. \textsc{Ascher}$^{4}$, B. \textsc{Bilgier}$^{6}$, L. \textsc{C{\'a}ceres}$^{7}$, R.~B. \textsc{Cakirli}$^{6}$, H. \textsc{Fujita}$^{3}$, E. \textsc{Ganio{\u{g}}lu}$^{6}$, M.\textsc{Gerbaux}$^{4}$, J. \textsc{Giovinazzo}$^{4}$, 
 S. \textsc{Gr{\'e}vy}$^{4}$,  O. \textsc{Kamalou}$^{7}$, H.~C. \textsc{Kozer}$^{6}$, L. \textsc{Kucuk}$^{6}$, T. \textsc{Kurtukian-Nieto}$^{4}$, 
 F. \textsc{Molina}$^{1,8}$, L. \textsc{Popescu}$^{9}$, A.~M. \textsc{Rogers}$^{10}$, G. \textsc{Susoy}$^{6}$, C. \textsc{Stodel}$^{7}$,
T. \textsc{Suzuki}$^{3}$, A. \textsc{Tamii}$^{3}$, J.~C. \textsc{Thomas}$^{7}$} 

\inst{$^{1}$Instituto de F{\'i}sica Corpuscular, CSIC-Universidad de Valencia, E-46071 Valencia, Spain\\
$^{2}$Department of Physics, Osaka University, Toyonaka, Osaka 560-0043, Japan\\
$^{3}$Research Center for Nuclear Physics, Osaka University, Ibaraki, Osaka 567-0047, Japan\\
$^{4}$Centre d'Etudes Nucl{\'e}aires de Bordeaux Gradignan, CNRS/IN2P3 - Universit{\'e} Bordeaux 1, 33175 Gradignan Cedex, France\\
$^{5}$Department of Physics, University of Surrey, Guildford GU2 7XH, Surrey, UK\\
$^{6}$Department of Physics, Istanbul University, Istanbul, 34134, Turkey\\
$^{7}$Grand Acc{\'e}l{\'e}rateur National d'Ions Lourds, BP 55027, F-14076 Caen, France\\

$^{8}$Comisi{\'o}n Chilena de Energ{\'i}a Nuclear, Casilla 188-D, Santiago, Chile\\
$^{9}$SCK.CEN, Boeretang 200, 2400 Mol, Belgium\\
$^{10}$Physics Division, Argonne National Laboratory, Argonne, Illinois 60439, USA}

\email{berta.rubio@ific.uv.es}

\recdate{September 30, 2014}

\abst{A very exotic decay mode at the proton drip-line, $\beta$-delayed $\gamma$-proton decay, has been observed in the $\beta$ decay of the $T_z$ = -2 nucleus $^{56}$Zn. Three $\gamma$-proton sequences have been observed following the $\beta$ decay.  The fragmentation of the IAS in $^{56}$Cu has also been observed for the first time. The results were reported in a recent publication. At the time of publication the authors were puzzled by the competition between proton and $\gamma$ decays from the main component of the IAS. 
Here we outline a possible explanation based on the nuclear structure properties of the three nuclei involved, namely $^{56}$Zn, $^{56}$Cu and $^{55}$Ni, close to the doubly magic nucleus $^{56}$Ni. From the fragmentation of the Fermi strength and the excitation energy of the two populated 0$^{+}$ states we could deduce the off-diagonal matrix element of the charge-dependent part of the Hamiltonian responsible for the mixing. These results are compared with the decay of  $^{55}$Cu with one proton less than $^{56}$Zn. For completeness we summarise the results already published.}

\kword{$\beta$-delayed $\gamma$-proton decay, $^{56}$Zn, Gamow-Teller transitions, Charge-exchange reactions, Isospin mixing, Isospin symmetry, fragmentation of the IAS, proton-rich nuclei.}

\begin{document}
\maketitle

\section{Introduction}

The study of the properties of nuclei far from stability is one of the main frontiers of modern nuclear physics. 
Among many possible observables for nuclear structure, the $\beta$-decay strengths provide important testing grounds for nuclear structure theories far from stability. The mechanism of $\beta$ decay is well understood and dominated by allowed Fermi (F) and Gamow-Teller (GT) transitions. A successful description of the nuclear structure of the states involved should provide good predictions for the corresponding transition strengths $B$(F) and $B$(GT).

We have studied the $\beta^{+}$ decay of the $T_{z} = -2$ $^{56}$Zn nucleus to the $T_{z} = -1 $ $^{56}$Cu nucleus and have reported our results in \cite{sonja_PRL}. We have observed competition between $\beta$-delayed proton and $\gamma$ emission in states well above the proton separation energy $S_{p}$. Moreover we observed $\beta$-delayed $\gamma$ rays that populate proton-unbound levels that subsequently decay by proton emission. This observation emphasizes the need for Ge detectors in order to determine the $B$(GT) correctly, even when the $\beta$-feeding is to  proton unbound levels.  In the present case, in order to determine $B$(GT) properly, the intensity of the proton transitions has to be corrected for the amount of indirect feeding coming from the $\gamma$ de-excitation. Although similar cases were suggested in the $sd$-shell \cite{Wrede2009,Pfutzner2012} and  observed in the decay of $^{32}$Ar \cite{Bhattacharya2008}, this exotic decay mode has been observed here for the first time in the $fp$ shell. Moreover how this new decay mode affects the determination of $B$(GT) has not been discussed prior to our paper \cite{sonja_PRL}. 

The proton decay of the IAS is usually isospin forbidden and consequently the slower $\gamma$ decay mode can compete even if the state is well above the $S_{p}$. Such cases have been observed before \cite{Dossat2007}. However the case of $^{56}$Zn decay is special, we have observed that its IAS in $^{56}$Cu is fragmented in  two 0$^{+}$states. These two states are an admixture of T=1 and T=2 configurations. The proton decay of the T=2 component is isospin forbidden, but the T=1 is not. In consequence,  the de-excitation of both 0$^{+}$ states  should be dominated by the fast proton decay. 
The observed competition between proton-and gamma-decay for the 3508 keV state (our sensitivity does not allow us to see this for the other state), makes this case both exotic and  interesting. We believe that this unusual behaviour is due to the structure of the nuclei involved and to the fact that we are close to the doubly-magic nucleus $^{56}$Ni. Here we summarise the results already published\cite{sonja_PRL} and also discuss the proton-$\gamma$  competition for the first time.  Another important difference with respect to the results presented in \cite{sonja_PRL} is that the mass excess of $^{56}$Cu has now been measured. A preliminary value for the  mass excess of $^{56}$Cu of -38530(90) keV has been reported 
\cite{china2014} which agrees with the value we used in   \cite{sonja_PRL} that was derived from systematics \cite{Audi2003}, rather than the value given in \cite{Audi2012}. Since the measured value is still preliminary, we have followed the same criteria as in 
 \cite{sonja_PRL} and used \cite{Audi2003} to calculate the $Q_{EC}^\#$ and $S_p$. The present experiment was motivated by a comparison with the mirror charge exchange (CE) reaction on $^{56}$Fe \cite{HFujita2010}. Indeed $\beta$ decay and CE studies are complementary and, assuming isospin symmetry, they can be combined to determine the absolute $B$(GT) values up to high excitation energies \cite{Taddeucci1987,Fujita2005,Fujita2011}. Hence precise determination of the $B$(GT) is important.

\section{Experiment}

The $\beta$-decay experiment was performed at the LISE3 facility of GANIL \cite{Mueller1991} in 2010, using a $^{58}$Ni$^{26+}$ primary beam with an average intensity of 3.7 e$\mu$A. This beam, accelerated to 74.5 MeV/nucleon, was fragmented on a 200 $\mu$m thick natural Ni target. The fragments were selected by the LISE3 separator and implanted into a 300 $\mu$m thick Double-Sided Silicon Strip Detector (DSSSD), surrounded by four EXOGAM Ge clovers for $\gamma$ detection. The DSSSD was used to detect both the implanted fragments and subsequent charged-particle decays ($\beta$s and protons). An implantation event was defined by simultaneous signals in both a silicon $\Delta E$ detector located upstream and the DSSSD. The implanted ions were identified by combining the energy loss signal in the $\Delta E$ detector and the Time-of-Flight (ToF)  difference between the cyclotron radio-frequency and the $\Delta E$ signal. Decay events were defined as giving a signal in the DSSSD and no coincident signal in the $\Delta E$ detector.

\section{Analysis and Results}

\begin{figure}[!t]
	\begin{minipage}{0.7\linewidth}
	  \hspace{12.0 mm}
		\includegraphics[height=1.\columnwidth, angle=90]{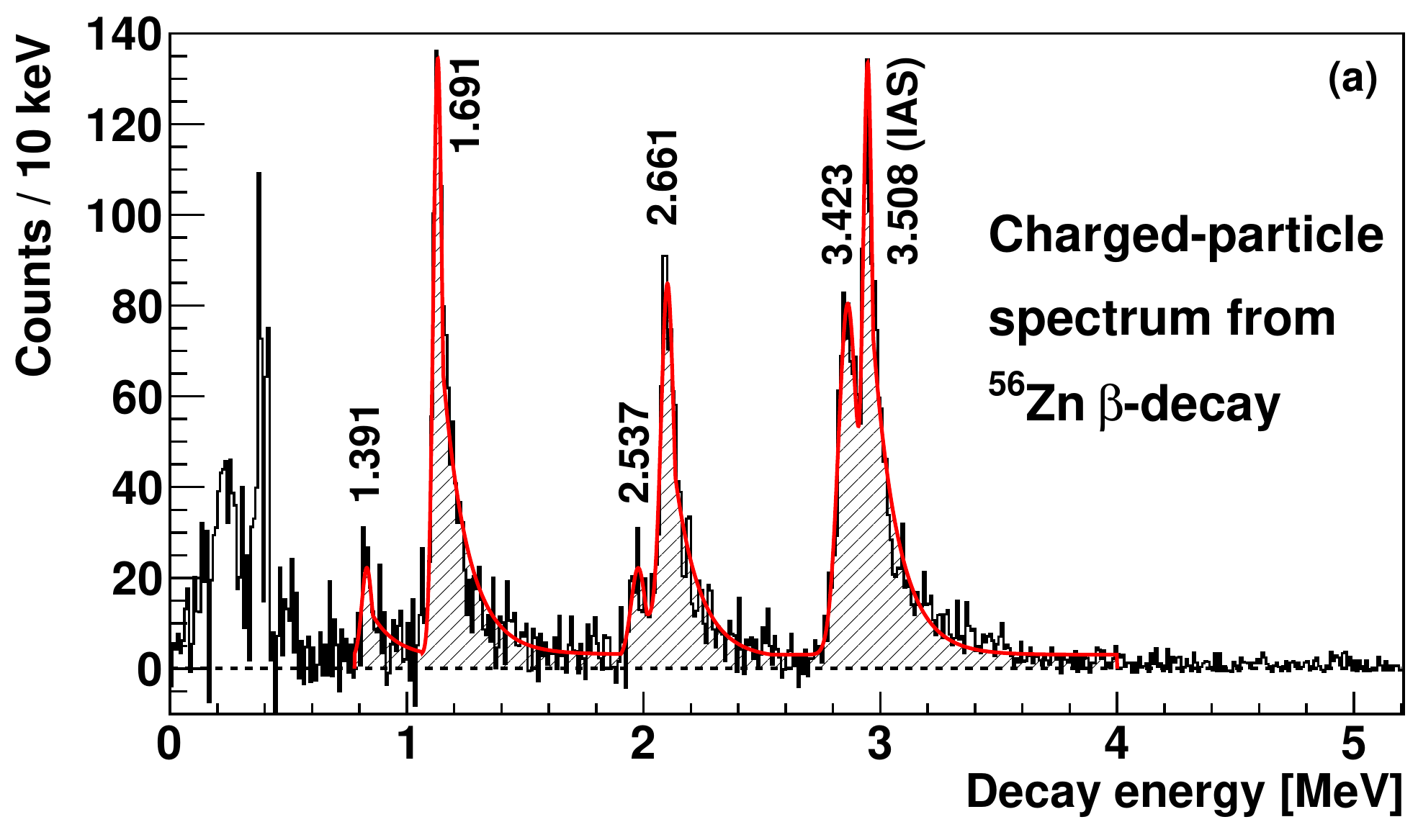}
	\end{minipage}
	\begin{minipage}{0.7\linewidth}
	  \hspace{12.0 mm}
         \includegraphics[height=1.\columnwidth, angle=90]{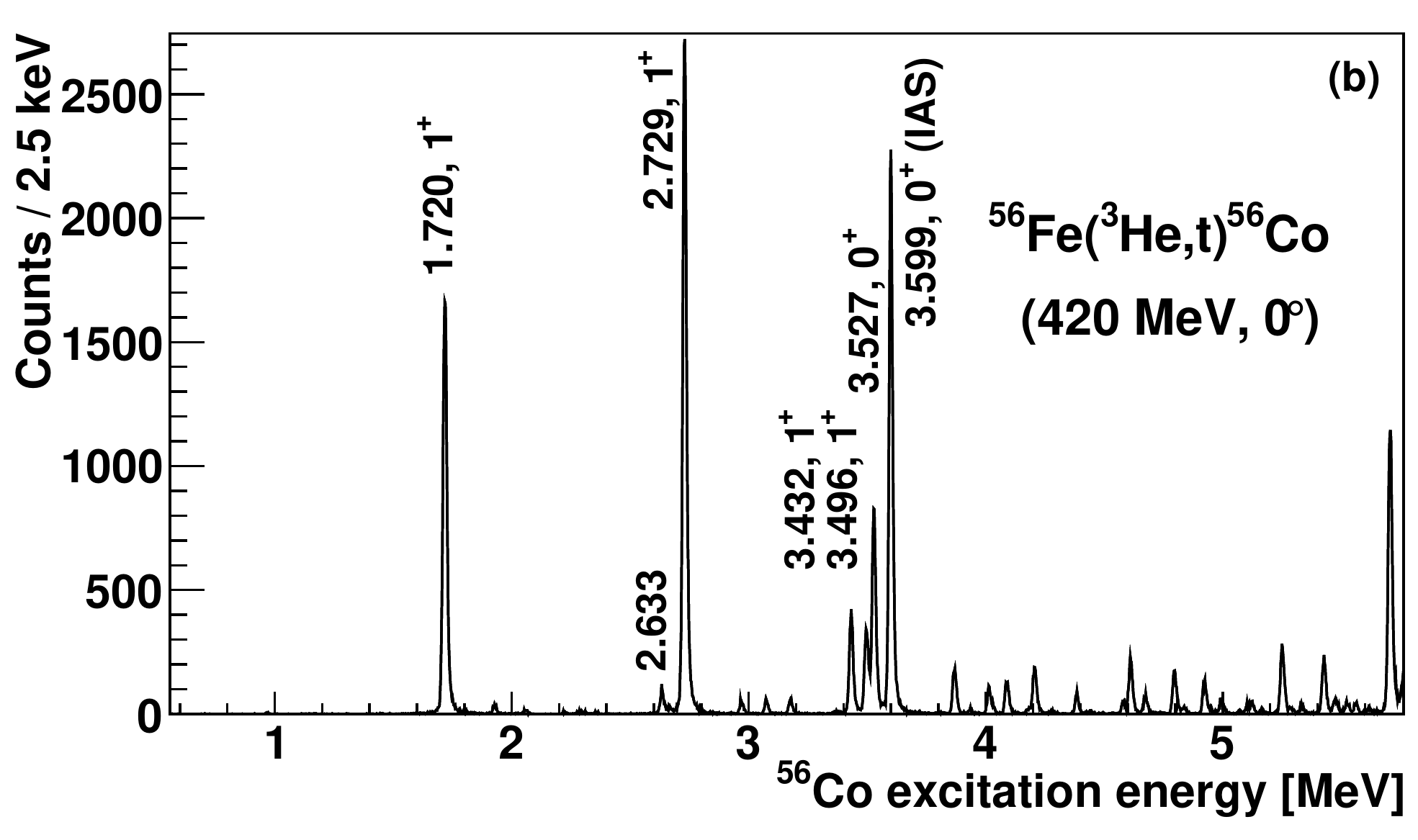}
	  \vspace{-5.0 mm}
	  \end{minipage}
		\caption{(a) Charged-particle spectrum measured in the DSSSD for decay events correlated with $^{56}$Zn implants. The peaks are labeled according to the corresponding excitation energies in $^{56}$Cu. (b) $^{56}$Fe($^{3}$He,$t$)$^{56}$Co reaction spectrum \cite{HFujita2010}. Peaks are labeled by the excitation energies in $^{56}$Co.}
		\label{p+CE-spectra}
	
	\vspace{-5.0 mm}
\end{figure}

The $^{56}$Zn ions were selected by setting gates off-line on the $\Delta E$-ToF matrix. The $\it{correlation~time}$ was defined as the time difference between a decay event in a given pixel of the DSSSD and any implantation signal that occurred before and after it in the same pixel that satisfied the conditions required to identify the nuclear species. The proton decays were selected by setting an energy threshold above 800 keV and looking for correlated $^{56}$Zn implants. 
The spectrum of the time correlations between the $^{56}$Zn implants and the protons was fitted with a function including the $\beta$ decay of $^{56}$Zn and a constant background. A half-life ($T_{1/2}$) of 32.9(8) ms was obtained for $^{56}$Zn \cite{sonja_PRL}, in agreement with \cite{Dossat2007}. Fig. \ref{p+CE-spectra}(a) shows the spectrum of charged-particles associated with $^{56}$Zn implants.

 Two kinds of state are expected to be populated in the $\beta$ decay of $^{56}$Zn to $^{56}$Cu: the $T$ = 2, $J^{\pi}$= 0$^{+}$ IAS, and a number of $T$ = 1, 1$^{+}$ states. From the comparison with the mirror nucleus $^{56}$Co, all of these states will lie above $S_p$ = 560(140) keV \cite{Audi2003} and will decay by protons. Indeed most of the strength in Fig. \ref{p+CE-spectra}(a) is interpreted as $\beta$-delayed proton decay to the $^{55}$Ni ground state. We attribute the broad bump below 800 keV to $\beta$ particles that are not in coincidence with protons. The proton peaks seen above 800 keV are labeled in terms of the excitation energies in $^{56}$Cu. The large uncertainty of $\pm$140 keV in the $^{56}$Cu level energies comes from the uncertainty in the estimated $S_p$. The proton decay of the IAS is identified as the peak at 3508 keV, as in \cite{Dossat2007}. Figure \ref{p+CE-spectra}(b) shows the triton spectrum from the mirror $T_{z} = +2 \rightarrow +1$, $\beta^{-}$-type CE reaction $^{56}$Fe($^{3}$He,$t$)$^{56}$Co recorded at RCNP 
 Osaka \cite{HFujita2010}. The corresponding $^{56}$Co states were determined with $\sim30$ keV resolution. Figs. \ref{p+CE-spectra}(a) and \ref{p+CE-spectra}(b) have been aligned in excitation energy in $^{56}$Cu and $^{56}$Co. There is a good correspondence between states in the mirror nuclei $^{56}$Cu and $^{56}$Co. 


\begin{figure}[!b]
  \vspace{-5.0 mm}
  \begin{minipage}{0.8\linewidth}
     \hspace{12.0 mm}
         \includegraphics[height=1.\columnwidth, angle=90]{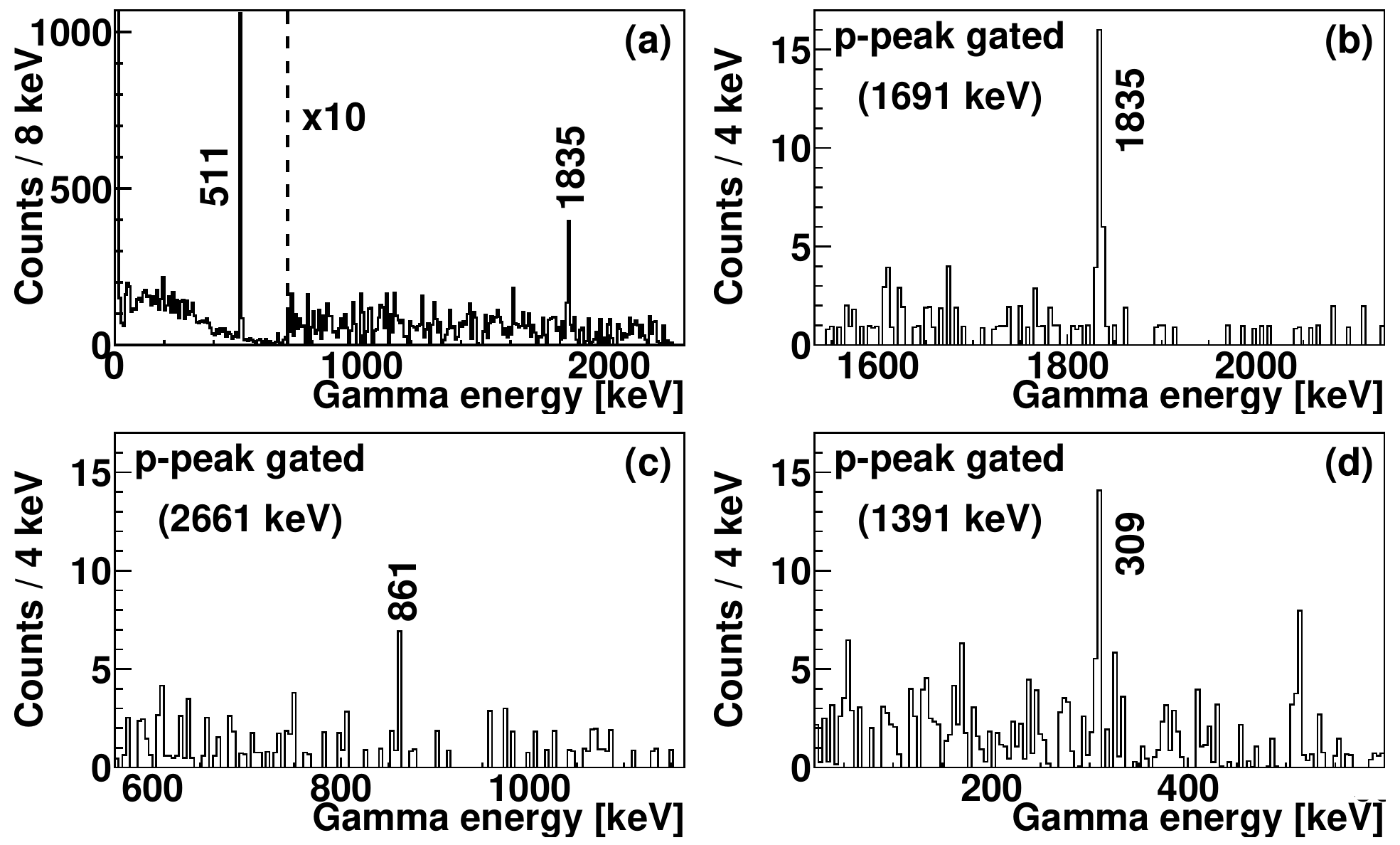}
	\caption{$\gamma$-ray spectra in coincidence with (a) all charged particles, and protons from the levels at (b) 1691, (c) 2661 and (d) 1391 keV.}
	\label{g-spectrum}
	\vspace{-5.0 mm}
	 \end{minipage}
\end{figure}

\begin{figure}[!t]
   \begin{minipage}{0.6\linewidth}
	  \hspace{12.0 mm}
         \includegraphics[height= 1. \columnwidth, angle=90]{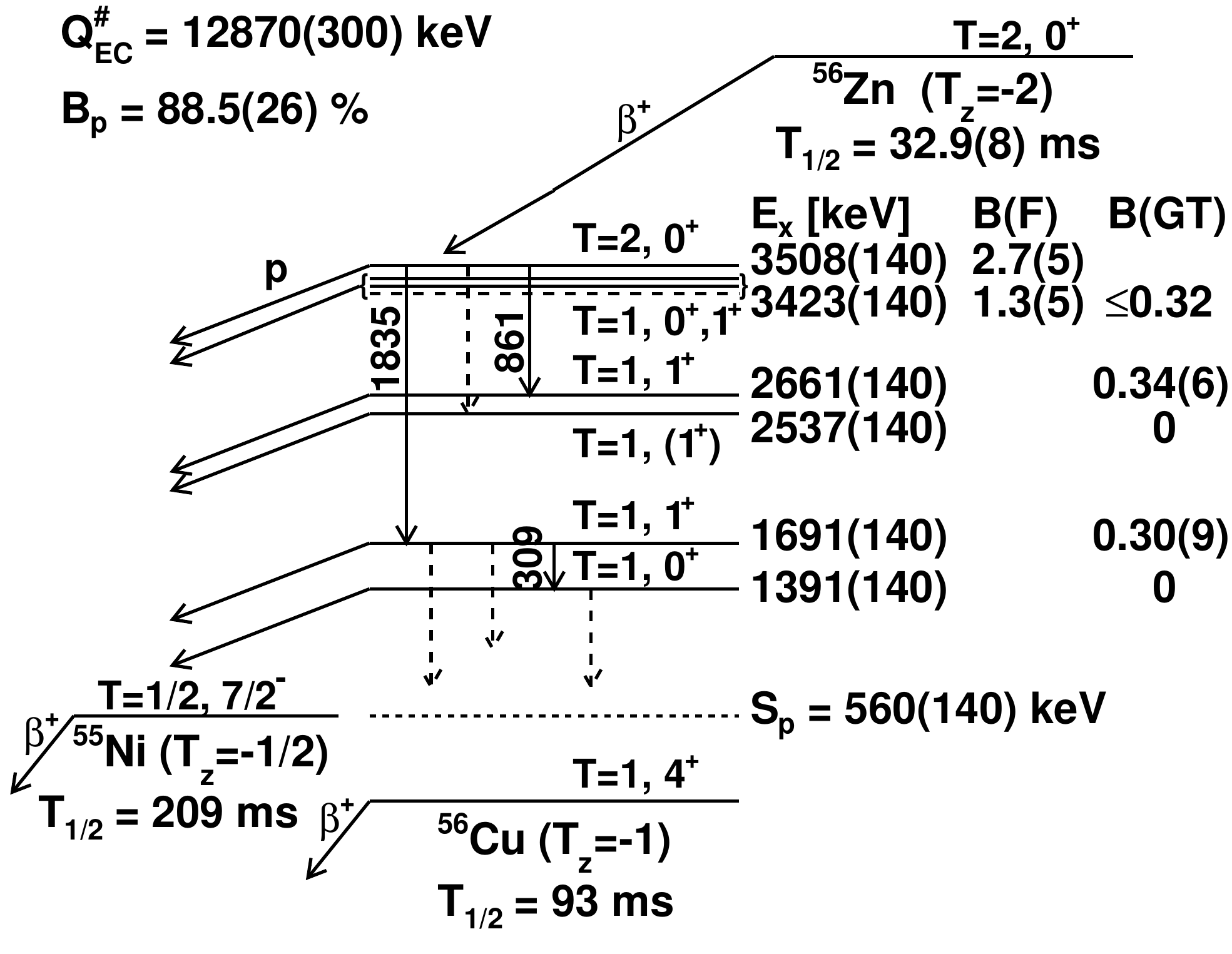}	
	\label{decay-scheme}
	\end{minipage}
	  \vspace{-5.0 mm}
  \caption{$^{56}$Zn decay scheme deduced from results of the present experiment. Observed proton or $\gamma$ decays are indicated by solid lines. Transitions corresponding to those seen in the mirror $^{56}$Co nucleus are shown by dashed lines. The error of 140 keV comes from the uncertainty in $S_p^\#$ (see text).} 
	\vspace{-5.0 mm}
\end{figure}

Imposing coincidence conditions on the total proton spectrum  or on the various proton peaks in Fig. \ref{p+CE-spectra}(a), 
three proton-$\gamma$-ray coincidence peaks were observed (Fig. \ref{g-spectrum}). 
All of our observations are discussed in \cite{sonja_PRL} and result in the $^{56}$Zn decay scheme shown in Fig. \ref{decay-scheme}.  The 831 keV proton transition could be placed as de-exciting  the 3508 or 1391 keV levels. The former is ruled out by the observed 831 keV proton-309 keV $\gamma$ coincidences (Fig. \ref{g-spectrum}(d)). Three cases of  $\beta$-delayed $\gamma$-proton emission have been established experimentally. Other cases of $\gamma$ decay from an IAS above $S_p$ have been observed in this mass region \cite{Dossat2007,Fujita2013}. The particular circumstance here is that the final level is also proton-unbound, consequently the $\beta$-delayed $\gamma$-proton decay has been observed for the first time in the $fp$-shell.
Table \ref{Table1} shows the center-of-mass energies of the proton and $\gamma$ peaks, and their intensities deduced from the areas of the peaks. For a proper determination of $B$(F) and $B$(GT), the $\beta$ feeding to each $^{56}$Cu level was estimated from the proton and $\gamma$ intensities, taking into account the amount of indirect feeding produced by the $\gamma$ de-excitation. In the case of the IAS we have assumed that the state decays in a similar way to its counterpart in the mirror therefore another $\gamma$-decay branch was added \cite{sonja_PRL}.
The measured $T_{1/2}$, $\beta$ feedings $I_\beta$ and $B_p$ together with the $Q_{EC}^\#$ and $S_p$ from \cite{Audi2003}   were used to determine the $B$(F) and $B$(GT) values shown in Table \ref{Table2} and in Fig. \ref{decay-scheme}.

The total Fermi transition strength has to be $|N-Z|$ = 4. The $^{56}$Cu IAS at 3508 keV has a Fermi strength $B$(F) = 2.7(5). The missing strength, 1.3(5), has to be hidden in the broad peak at 3423 keV. This is a confirmation that the $^{56}$Cu IAS is fragmented and thus most of the feeding to the 3423 keV level (assuming it contains two or three unresolved levels) corresponds to the Fermi transition and the remainder to the GT transition. The isospin impurity $\alpha^2$ (defined as in \cite{HFujita2010}) and the off-diagonal matrix element of the charge-dependent part of the Hamiltonian $\left\langle{H_{c}}\right\rangle$, responsible for the isospin mixing of the 3508 keV IAS (0$^{+}$, $T=2$) and the 0$^{+}$ part of the 3423 keV level ($T=1$), were derived  assuming two-level mixing. For $^{56}$Cu it was found that $\left\langle{H_{c}}\right\rangle$ = 40(23) keV and $\alpha^2$ = 33(10)\%, similar to the values obtained in the mirror $^{56}$Co \cite{HFujita2010}, $\left\langle{H_{c}}\right\rangle$ = 32.3(5) keV and $\alpha^2$ = 28(1)\%. It is interesting to compare our results on the decay of $^{56}$Zn with $T_{z} = -2$ with the investigation on the decay of $^{55}$Cu with $T_{z} = -3/2$, only one proton apart. In a recent article \cite{Tripathi2013} a strong fragmentation of the IAS was observed. We note that in \cite{Tripathi2013} the states could be fed by both F and GT transitions, although the GT component is small. Similarly here the second 0$^{+}$ state (fed by F) could not be separated from a 1$^{+}$ state (fed by GT) although the feeding to the 1$^{+}$ state is small. In the $^{55}$Ni case, 
 $\left\langle{H_{c}}\right\rangle$ = 9(1) keV and $\alpha^2$ = 27(7)\%, also in accord with observations in the mirror. In both cases the mixing matrix element is small, but it seems to increase when moving away from N = Z. 
\\


\begin{figure}[!t]
 \vspace{-2.0 mm}
	\includegraphics[height=0.45\columnwidth]{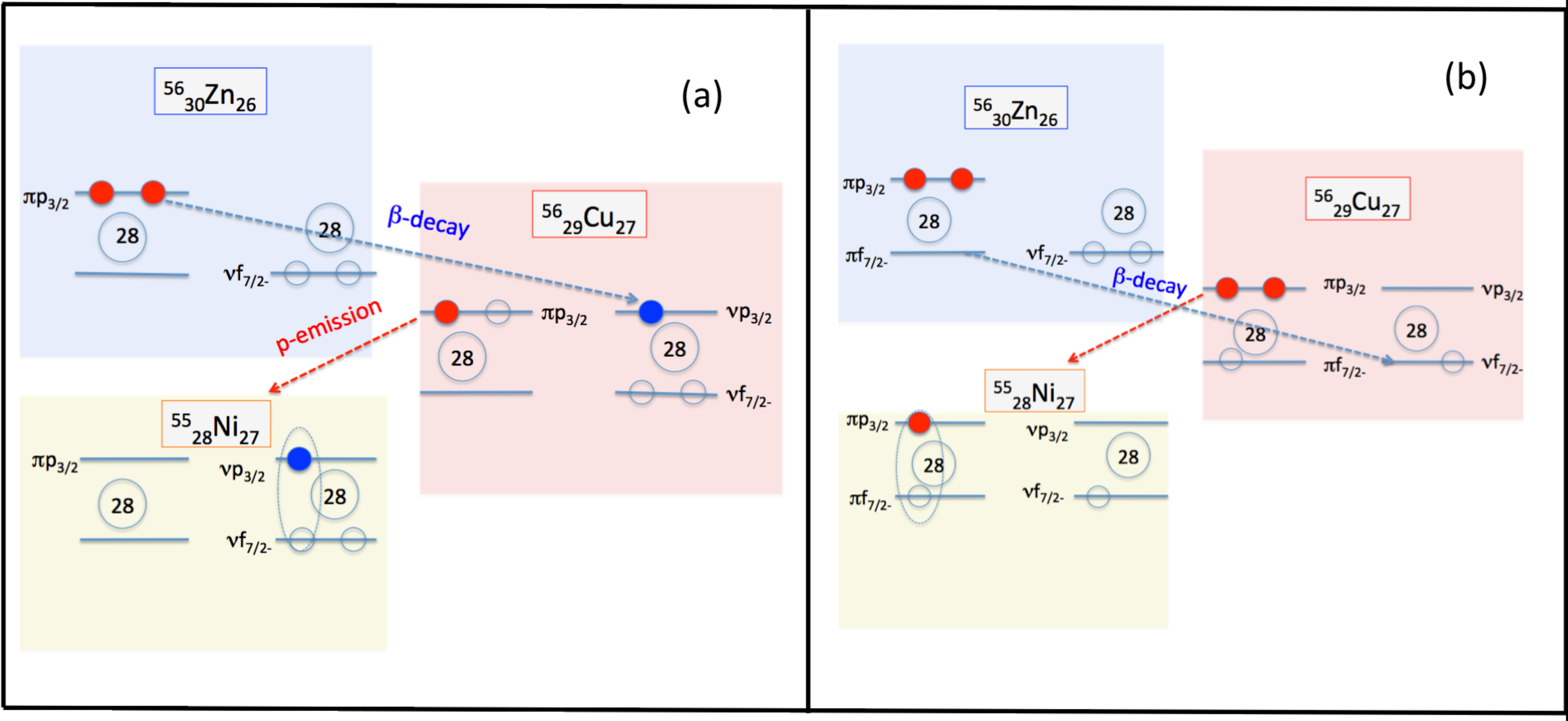}
	 \vspace{-3.0 mm}
  \caption{A schematic diagram showing two possible explanations for the observed decay of the $^{56}$Zn IAS (see text).}
	\label{explanation}
\vspace{-8.0 mm}
\end{figure}

\vspace{-5mm}
\begin{table}[tbh]
\parbox{0.45\linewidth}{
  \caption{Center-of-Mass proton energies, $\gamma$-ray energies, and their intensities (normalized to 100 decays) for the decay of $^{56}$Zn. *IAS.}
	\label{Table1}
		\centering
	  \begin{tabular}{lcccc}
	  \hline

		  $E_p$(keV) & $I_p$(\%) & $E_{\gamma}$(keV) & $I_{\gamma}$(\%)\\ \hline
		  2948(10)* & 18.8(10) & 1834.5(10) & 16.3(49)\\
		  2863(10) & 21.2(10) & 861.2(10) & 2.9(10)\\
		  2101(10) & 17.1(9) & 309.0(10) & - \\
		  1977(10) & 4.6(8) & & \\
		  1131(10) & 23.8(11) & & \\
		  831(10) & 3.0(4) & & \\
		  \hline
	  \end{tabular}
}
\hfill
\parbox{0.45\linewidth}{
	\caption{$\beta$ feedings for Fermi and Gamow Teller transition strengths to levels in  $^{56}$Cu  in the $\beta^{+}$ decay of $^{56}$Zn calculated using \cite{Audi2003}. *IAS.}
	\label{Table2}
	\centering
	  \begin{tabular}{lcccc}
	   \hline

		  $I_\beta$(\%) & $E$(keV) & $B$(F)$$ & $B$(GT)$$ \\ \hline
		  43(5) & 3508(140)* & 2.7(5) & \\
		  21(1) & 3423(140) & 1.3(5) & $\leq$0.32 \\
		  14(1) & 2661(140) & & 0.34(6) \\
		  0 & 2537(140) & & 0 \\
		  22(6) & 1691(140) & & 0.30(9) \\
		  0 & 1391(140) & & 0 \\
		   \hline
	  \end{tabular}
}
 \vspace{-5mm}	  
\end{table}
An interesting and puzzling open question is, considering that the isospin mixing is quite large in  $^{56}$Cu, why we are still observing the $\gamma$ de-excitation from the IAS when the partially-allowed proton decay is, in principle, much faster. The mirror level in
$^{56}$Co lies at 3600 keV. It has $t_{1/2}=2 \times 10^{-14}$ s, and decays only by $\gamma$ de-excitation, with the strongest of the three branches having an energy similar to that of the corresponding gamma-ray observed in the decay of the $^{56}$Cu IAS. This allows us to estimate the 3508 keV  partial $\gamma$ half-life.  The proton half-life of the IAS in $^{56}$Cu is estimated to be of the order of 10$^{-18}$ s, i.e. 10$^{4}$ times faster. One possible answer to this dilemma lies in the nuclear structure of the levels involved. In Fig. \ref{explanation} we sketch the way the decay may proceed. We start with the assumption that the  $^{56}$Zn ground state ($gs$) has the two extra protons above the Z$=$28 gap in  the $\pi p_{3/2}$ orbital and the two neutron holes in the  $\nu f_{7/2}$. In Fig. \ref{explanation}a the decay proceeds by transforming a proton in the  $\pi p_{3/2}$ into a neutron in  $\nu p_{3/2}$.
The subsequent proton decay  will most probably remove the unpaired proton  from the $\pi p_{3/2}$ orbital.
As shown in the figure, the final state in $^{55}$Ni, will consist of a neutron particle-hole excitation across the N=28 gap on top of the neutron hole in the 
$\nu f_{7/2}$ orbital which is the most probable configuration for the $gs$ in $^{55}$Ni. This neutron particle-hole excitation will lie at 4 to 5 MeV excitation energy and is therefore outside the available energy gap for the protons of $\sim$3 MeV (3508-560 keV, see Fig. \ref{decay-scheme}).  In Fig. \ref{explanation}b we assume the same $gs$ for $^{56}$Zn but in this case the decay proceeds through the transformation of a proton from the $\pi f_{7/2}$ level into one of the two available holes in the $\nu f_{7/2}$ shell.
The proton decay will probably remove one of the two available protons in the $\pi p_{3/2}$ orbital in $^{56}$Cu giving rise to a final proton particle-hole excitation on top of the $\nu f_{7/2}$ hole  in the $^{55}$Ni $gs$ which should again lie at 4 to 5 MeV above the $gs$ in $^{55}$Ni.
We note that the same argument applies if we start by assuming that the two protons occupy the $\pi f_{5/2}$ or the $\pi p_{1/2}$ orbitals (moreover, if the proton decay originates from the $\pi f_{7/2}$ the situation is even less favourable since the final state will involve two particle-hole excitations across the the N and Z $=$ 28 gaps). In summary, the  proton decay of the IAS of $^{56}$Zn into $^{55}$Ni is strongly hindered from the nuclear structure point of view. This may explain why the 
$\gamma$-decay can compete with it. Shell model calculations to confirm these ideas are in progress.



\section{Acknowledgments}
This work was supported by the Spanish MICINN grants FPA2008-06419-C02-01, FPA2011-24553; CPAN Consolider-Ingenio 2010 Programme CSD2007-00042; MEXT, Japan 18540270 and 22540310; Japan-Spain coll. program of JSPS and CSIC; Istanbul University Scientific Research Projects, Num. 5808; UK (STFC) Grant No. ST/F012012/1; Region of Aquitaine. R.B.C. acknowledges support by the Alexander von Humboldt foundation and the Max-Planck-Partner Group. We acknowledge the EXOGAM collaboration for the use of their clover detectors. We thank I. Hamamoto for valuable help in the interpretation of our observations.


\end{document}